\documentclass[12pt]{iopart}
\usepackage{datetime}

\usepackage{iopams}
\usepackage{epsf}
\usepackage{epsfig}
\usepackage{color}
\usepackage{psfrag}
\usepackage{verbatim}
\usepackage{setstack}
\usepackage{amsopn}
\usepackage{appendix}
\usepackage{graphicx}
\usepackage[small]{caption}
\newcommand{\be}{\begin{eqnarray}}
\newcommand{\ee}{\end{eqnarray}}

\begin{document}

\unitlength = 1mm
\eqnobysec

\title{Six--vertex model with domain wall boundary conditions in the
  Bethe--Peierls approximation}
\author{L. F. Cugliandolo$^1$, G. Gonnella$^2$ and A. Pelizzola$^{3,4,5}$}

\address{$^1$
Sorbonne Universit\'es, Universit\'e Pierre et Marie Curie - Paris 6, 
Laboratoire de Physique Th\'eorique et Hautes Energies,
4 Place Jussieu, 
75252 Paris Cedex 05, France\\
$^2$ Dipartimento di Fisica, Universit\`a di Bari {\rm and} 
INFN, Sezione di Bari, 
via Amendola 173, Bari, I-70126,
Italy\\
$^3$ Dipartimento di Scienza Applicata e Tecnologia, CNISM and
  Center for Computational Studies, Politecnico di Torino,
Corso Duca degli Abruzzi 24, I--10129 Torino, Italy \\
$^4$ INFN, Sezione di Torino, 
via Pietro Giuria 1, I-10125 Torino,
Italy \\
$^5$ Human Genetics Foundation, HuGeF, 
Via Nizza 52, I-10126 Torino,
Italy}

\begin{abstract}
We use the Bethe--Peierls method
combined with the belief propagation algorithm to study the arctic curves in the six vertex model 
on a square lattice with domain-wall boundary conditions, and the six vertex model on a rectangular lattice with 
partial domain-wall boundary conditions. We show that this rather simple approximation 
yields results that are remarkably close to the exact ones when these are known, and allows one to 
estimate the location of the phase boundaries with relative little effort in cases in which exact results are not
available.
\end{abstract}

\noindent
\today \hspace{0.5cm} 

\tableofcontents

\section{Introduction}
\label{sec:intro}

Usually, boundary conditions have no effect on the equilibrium properties of statistical models in the thermodynamic limit.
Still, boundary conditions may affect the equilibrium behaviour of strongly constrained models and, in particular, they 
may impose the macroscopic
separation of phases. In such cases, the boundary between an internal spatial region, in which the 
configuration is typical of equilibrium with the bulk parameters,
and an external spatial region, frozen by the boundary conditions is named the {\it arctic curve}.  
The external region is called  
 {\it arctic} (for frozen) and the internal region is called {\it temperate} (for fluctuating). 
The arctic curve first appeared in 
the study of domino tilings of Aztec diamonds~\cite{Elkies1992a,Elkies1992b,Cohn1996,Jockusch1998}, then in lozenge tilings of large hexagons~\cite{Cohn1998,Borodin2010}, 
and later in more general dimer~\cite{Kenyon} and vertex models~\cite{Cohn1996,Jockusch1998,ColomoPronko-proc,ColomoPronko2010c,ColomoPronko2010b,ColomoPronkoZinnJustin2010,ColomoNoferiniPronko}. In these systems 
phase separation exists for a wide choice of fixed boundary conditions.

The six vertex model was first introduced to model 
ferroelectricity~\cite{Lieb1967a,Lieb1967b,Lieb1967c,Wuphasetransitions,BaxterBook},
and it has links with the models mentioned above.
It is commonly defined on a square lattice with $N\times N$ vertices.
(We will consider here the extension to rectangular systems with $N\times M$ vertices as well.)
Arrows are placed along the links, with two possible orientations for each. The six vertex rule imposes
that two arrows should point in and two should point out of each vertex. Energies and, consequently, 
statistical weights are assigned to each vertex. If complete arrow reversal symmetry is assumed, 
three parameters, $a$, $b$, $c$, characterise these weights, with the two first ones associated to local 
ferroelectric order and the last one linked to antiferroelectric order (see Fig.~\ref{fig:VertexWeights}).

The solution to the most general six vertex model with {\it periodic boundary conditions},  
in the form of its bulk free-energy density,  was given by Sutherland~\cite{Sutherland1967}
who extended Lieb's treatment with the Bethe {\it Ansatz} technique to generic values of the
vertex weights.
The value taken by the following parameter (see Fig.~\ref{fig:VertexWeights}) 
\begin{equation}
\Delta = \frac{a^2+b^2-c^2}{2ab}
\end{equation}
distinguishes the three phases of the model: 
for $\Delta>1$ the equilibrium state has ferroelectric (FE) order, for $\Delta < -1$ it has antiferroelectric (AF) order, 
and for $|\Delta|<1$ it is disordered (D).  The D-FE phase transition is first-order while the D-AF transition is characterized by 
an essential singularity reminiscent of the Kosterlitz-Thouless type. 

The six vertex model with {\it domain wall boundary conditions} was introduced in~\cite{Korepin1982} to 
study correlation functions in exactly solvable models~\cite{KorepinBogoliubovIzergin}. In a few words, domain-wall boundary conditions correspond to 
all arrows on the bottom and top boundaries being incoming while all arrows on the left and right boundaries being outgoing, 
see Fig.~\ref{fig:dwbc} (or {\it vice versa}).
More recently, the interest in this model has been pushed by its connection with algebraic combinatorics
and the enumeration of alternating sign matrices. 
The partition function satisfies a recurrence relation~\cite{Izergin87,Korepin1982}, that leads to a determinant formula~\cite{Izergin87}, 
exploited by Korepin and P.  Zinn-Justin (see also~\cite{Kuperberg96}) to derive exact expressions for 
the free-energy densities in the disordered and 
ferroelectric phases~\cite{KorepinZinnJustin00}. The antiferroelectric free-energy density was obtained in~\cite{ZinnJustin2000} by using 
a matrix model representation of the determinant. Although the phase diagram remains unaltered,
and still given by the case with $|\Delta |=1$,  the free-energy densities  in the disordered and antiferromagnetic 
phases, are different from the ones found with periodic boundary conditions,
even in the thermodynamic limit.  Curiously enough, the free-energies take much simpler expressions as 
functions of the parameters $a, b, c$ under the domain-wall boundary conditions.  The D-FE transition becomes continuous
while the D-AF remains of infinite order.

The six vertex model can be mapped onto the problem of domino tilings on Aztec diamonds~\cite{BaxterBook,Wuphasetransitions} 
for a particular choice of vertex weights~\cite{Korepin1982}
and, in consequence, it also admits an arctic curve separating an external finite density boundary with ferroelectric order from
an internal region that is either disordered or antiferroelectrically ordered  for $|\Delta|<1$ or $\Delta <-1$, respectively.
On the free-fermion point ($\Delta=0$) with $a=b$  the arctic curve is a circle~\cite{Cohn1996,Jockusch1998}.
On the less restrictive free-fermion cases with $a\neq b$ the arctic curves are ellipses~\cite{ColomoPronko-proc}.
Numerical evidence for the existence of an arctic curve for general values of the parameters
was presented in~\cite{Syljuasen2004,Allison2005}.
Various analytic, though not yet proven to be exact (except for $\Delta = 0$ where the derivation is rigorous),
methods have been employed to determine the arctic curves in the disordered
phase with $\Delta\neq 0$~\cite{ColomoPronko2010c,ColomoPronko2010b} and even in the antiferroelectric 
phase~\cite{ColomoPronkoZinnJustin2010}.

As mentioned above, very powerful and interesting analytic methods allow one to 
understand the statics of the six vertex model, and the phase separation
induced by special boundary conditions. A summary of modern methods applied to this problem can be found 
in~\cite{Reshetikhin} and a review on arctic curves in the six vertex model is given in~\cite{ColomoNoferiniPronko}. 
Still, as soon as one lifts the integrability conditions 
satisfied by the six (and eight) vertex model, these techniques are no longer useful. Approximate methods, such as 
the Bethe--Peierls approximation~\cite{Bethe35} and its modern versions, like the cavity method and the belief propagation 
algorithm~\cite{Pearl,Yedidia2003,CVMreview,MezardMontanari}, can then be of great help to obtain the phaser diagram and 
equilibrium properties of generic vertex models. This method was applied in~\cite{Foini2013,Levis2013a}
to analyse the sixteen vertex model with parameters close to the ones of artificial spin-ice samples~\cite{Heyderman2013,NisoliMoessnerSchiffer2013}.
Surprisingly enough, the method  gave very accurate, sometimes even exact, results when applied to the 
integrable six and eight vertex cases.
Moreover, cluster generalizations have been used to obtain phase diagrams and thermodynamical properties of 
vertex models with a much larger state space \cite{CGP1996a,CGP1996b,CGP1997,CGJP1997,CGP2000,CGP2012}, both in two and 
three dimensions.

The work in~\cite{Foini2013,Levis2013a} concerned homogeneous systems (except for distinguishing between two different 
sublattices, when necessary) with periodic boundary conditions. Here we aim at extending this work to inhomogeneous systems, 
making it possible to describe the phase separation phenomenon induced by domain--wall boundary conditions. This will imply an 
increase in the computational complexity by a factor $N \times M$ which, even in the large $N$ and $M$ limit, can be dealt 
with thanks to the belief propagation (BP) algorithm, which is particularly efficient at finding minima of a Bethe--Peierls free-energy 
even in inhomogeneous systems. As an example of results that we obtain  in this work,  for which there are no rigourous predictions,  there is the behavior 
of the interface separating the disordered region from the AF phase  
in the case of parameters corresponding to the AF bulk phase.

We end this introductory section by stating that domain wall boundary conditions could be easily imposed experimentally in artificial spin-ice 
samples~\cite{Heyderman2013,NisoliMoessnerSchiffer2013}. A possible way to do this would be to fabricate an array where the edge 
islands are different in some way (larger, or from a different magnetic material) in order to make them more stable. 
Modern visualization experimental methods~\cite{PhysRevB.83.174431,PhysRevB.78.144402,PhysRevLett.111.057204,Ladak10} 
should then allow one to see the phase separating curves in the lab.

The paper is organised as follows. In Sec.~\ref{sec:model} we recall the definition of the six vertex model and 
the properties that are of  interest to our work. Section~\ref{sec:method} is devoted to a short description of the 
belief propagation method as applied to the six vertex model with domain wall boundary conditions.
In Sec.~\ref{sec:results} we present our results. 
  In Sec.~\ref{sec:conclusions}
we discuss several lines for future work.

\section{The six vertex model and its arctic curve}
\label{sec:model}
 
\subsection{Definition, phases and free-energies}

The six vertices defining the model  together with their statistical weights are shown in Fig.~\ref{fig:VertexWeights}.

\setlength\unitlength{1cm}
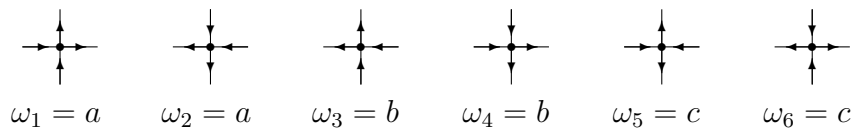
\begin{figure}[h]
\begin{center}
\begin{picture}(12,2)(0,-.75)
\put(1,.5){\circle*{.1}}
\put(1,0){\line(0,1){1}}
\put(.5,.5){\line(1,0){1}}
\put(1,.5){\vector(1,0){.35}}
\put(1,.5){\vector(0,1){.35}}
\put(.5,.5){\vector(1,0){.35}}
\put(1,0){\vector(0,1){.35}}
\put(.35,-.5){$\omega_1 = a$}
\put(3,.5){\circle*{.1}}
\put(3,0){\line(0,1){1}}
\put(2.5,.5){\line(1,0){1}}
\put(3.5,.5){\vector(-1,0){.35}}
\put(3,1){\vector(0,-1){.35}}
\put(3,.5){\vector(-1,0){.35}}
\put(3,.5){\vector(0,-1){.35}}
\put(2.35,-.5){$\omega_2 = a$}
\put(5,.5){\circle*{.1}}
\put(5,0){\line(0,1){1}}
\put(4.5,.5){\line(1,0){1}}
\put(5.5,.5){\vector(-1,0){.35}}
\put(5,.5){\vector(0,1){.35}}
\put(5,.5){\vector(-1,0){.35}}
\put(5,0){\vector(0,1){.35}}
\put(4.35,-.5){$\omega_3 = b$}
\put(7,.5){\circle*{.1}}
\put(7,0){\line(0,1){1}}
\put(6.5,.5){\line(1,0){1}}
\put(7,.5){\vector(1,0){.35}}
\put(7,1){\vector(0,-1){.35}}
\put(6.5,.5){\vector(1,0){.35}}
\put(7,.5){\vector(0,-1){.35}}
\put(6.35,-.5){$\omega_4 = b$}
\put(9,.5){\circle*{.1}}
\put(9,0){\line(0,1){1}}
\put(8.5,.5){\line(1,0){1}}
\put(9.5,.5){\vector(-1,0){.35}}
\put(9,.5){\vector(0,1){.35}}
\put(8.5,.5){\vector(1,0){.35}}
\put(9,.5){\vector(0,-1){.35}}
\put(8.35,-.5){$\omega_5 = c$}
\put(11,.5){\circle*{.1}}
\put(11,0){\line(0,1){1}}
\put(10.5,.5){\line(1,0){1}}
\put(11,.5){\vector(1,0){.35}}
\put(11,1){\vector(0,-1){.35}}
\put(11,.5){\vector(-1,0){.35}}
\put(11,0){\vector(0,1){.35}}
\put(10.35,-.5){$\omega_6 = c$}
\end{picture}
\end{center}
\caption{The six vertices in the six vertex model and their Boltzmann weights, $\omega_1, \dots, \omega_6$.}
\label{fig:VertexWeights}
\end{figure}

An alternative parametrisation of the Boltzmann weights  is 
\begin{equation}
a = \sin(\lambda+\eta) 
\qquad\;\;\;
b = \sin(\lambda-\eta)
\qquad\;\;\;
c=\sin 2\eta 
\end{equation}
with $\lambda$ the `rapidity' variable and $\eta$ the `crossing' parameter.
The parameter $\Delta=\cos2\eta$  serves to locate the phase transitions.

In the disordered (D) 
phase $|\Delta|=|\cos2\eta| <1$ and the parameters $\lambda$ and $\eta$ are constrained to  $\eta\leq \lambda \leq \pi - \eta$ and 
$0 < \eta < \pi/2$. In the free-fermion case $\Delta=0$ and $\eta=\pi/4$.
In the free-fermion and symmetric $a=b$ case, $\Delta=0$, $\eta=\pi/4$ and $\lambda=\pi/2$.
In the spin-ice case $a=b=c$ implies $\Delta=1/2$, $\eta = \pi/6$ and $\lambda = \pi/2$.

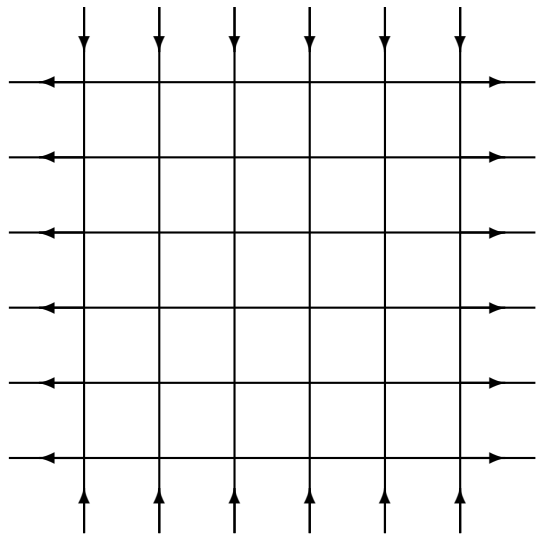
\begin{figure}[h]
\begin{center}
\begin{picture}(7,7)(-1,-1)
\thicklines
\multiput(0,-1)(1,0){6}{\line(0,1){7}}
\multiput(0,-1)(1,0){6}{\vector(0,1){.6}}
\multiput(0,6)(1,0){6}{\vector(0,-1){.6}}
\multiput(-1,0)(0,1){6}{\line(1,0){7}}
\multiput(0,0)(0,1){6}{\vector(-1,0){.6}}
\multiput(5,0)(0,1){6}{\vector(1,0){.6}}
\end{picture}
\end{center}
\caption{An example of domain wall boundary conditions.
}
\label{fig:dwbc}
\end{figure}

Let us first focus on the square lattice model.
The domain-wall boundary conditions (DWBC) are sketched in Fig.~\ref{fig:dwbc}.
The free-energy density per site in the disordered phase with these boundary conditions   
is~\cite{Korepin1982,KorepinZinnJustin00}
\begin{equation}
f^D_{DW} = - \ln \left( \frac{\alpha \sin (\lambda+\eta) \sin (\lambda-\eta)}{\sin[\alpha(\lambda-\eta)]} \right)
\qquad
\mbox{with}
\qquad
\alpha = \frac{\pi}{\pi-2\eta}
\; . 
\end{equation}
When compared to the (much more involved) expression for periodic boundary conditions~\cite{Sutherland1967} one finds
$f^D_{DW} > f^D_{P}$. The extension of this analysis to the 
antiferroelectric (AF) and ferroelectric (FE) phases shows that $f^{AF}_{DW}>f^{AF}_{P}$ in the AF phase~\cite{ZinnJustin2000} while 
  $f^{FE}_{DW} = f^{FE}_{P}$ in the FE one~\cite{Korepin1982,KorepinZinnJustin00}.  The macroscopic phase separation 
is intimately linked to  the difference between the  free-energies for different boundary conditions. The D-FE phase transition 
is continuous and the D-AF transition is of infinite order under DWBC.

\subsection{The arctic curve}

The arctic curve(s) delimiting the spatial regions with different ordering properties 
is (are) defined in the scaling limit in which the number of lines in each direction, 
$N$ on a $N x N$ square lattice, tends to infinity while the lattice spacing, $\delta$, vanishes
keeping the two linear size of the lattice fixed.

In the case in which the bulk parameters select a disordered state, 
the phase separation induces five spatial regions in the sample: four external ferroelectric ones,
close to the four corners 
and an internal one, that is the disordered region  $D$.

In the case in which the bulk parameters select the AF state, the 
phase separation occurs in two steps, in the sense that an intermediate 
disordered spatial region separates the external FE region fixed by the 
boundaries, and the internal AF region fixed by the bulk parameters. There is therefore one
arctic curve and one temperate (between D and AF regions) curve in this problem.
  
Some analytic expressions for the interface between frozen and temperate regions in the 
six vertex model on a square lattice are known and we summarise them below.
  
\subsubsection{Boundary between the FE and the D phase.}
  
Colomo and Pronko used boundary correlation functions to get a closed expression for the coordinates
of the {\it contact points} between the arctic curve and the system's boundaries~\cite{ColomoPronko2010b}
\begin{equation}
\kappa = \frac{\alpha \cot[\alpha(\lambda-\eta)] - \cot(\lambda+\eta)}{\cot(\lambda-\eta) -\cot(\lambda+\eta)}
\end{equation}
for generic $|\Delta|<1$. In particular, for $\lambda=\pi/2$ one finds $\kappa=1/2$ for all $\eta$.

The form of the arctic curve linking these points is highly non-trivial and for generic $\eta$ non-algebraic. 
Some special cases have been known for more than a decade.

Jockusch et al~\cite{Jockusch1998} proved the arctic circle theorem for the free-fermion case $\Delta=0$ with  $a=b$
in which the arctic curve is just a circle,
\begin{equation}
\left( x-\frac12 \right)^2 + \left(y-\frac12 \right)^2 = \frac14
\; . 
\end{equation} 

In the free fermion case, $\Delta=0$ and $\eta=\pi/4$, the arctic curve can be computed exactly for $a\neq  b$ and it is an 
ellipse tangent to the four sides of the square
\begin{equation}
(1+t^2) (x+y -1)^2 + t^2 (1+t^2) (x-y)^2 = t^2
\end{equation}
with 
\begin{equation}
t = \frac{b}{a} = \tanh\left(\lambda - \frac{\pi}{4}\right)
\; . 
\end{equation}
For $t=1$ one recovers the article circle.

With the emptiness correlation function, and a conjecture on its behavior, 
Colomo and Pronko obtained an implicit expression for the arctic curve in the general $\Delta\neq 0$ 
case~\cite{ColomoPronko2010c,ColomoPronko2010b,ColomoSportiello}.
In section 6.3 in~\cite{ColomoPronko2010b}, eqs.~(6.17)-(6.19) present a parametric representation of this curve. 
In general, the curve is a transcendental function. For special values of $\Delta$, that correspond to what are called the 
roots-of-unity in the quantum group context, the curve becomes algebraic~\cite{ColomoPronko2010b,ColomoNoferiniPronko}.
In particular, at the spin-ice point ($a=b=c$ and $\Delta=1/2$) the arctic curve is given by the portion of the ellipse
\begin{equation}
(x+y-2)^2 + 3 (x-y)^2 - 3 =0
\end{equation}
in which $x, y \in[0,1/2]$. The contact points in the lower left quadrant  are $(1/2, 0)$ and $(0,1/2)$, in other words, $\kappa=1/2$.

\subsubsection{Boundary between the FE and AF phases.}

Colomo, Pronko and Zinn-Justin~\cite{ColomoPronkoZinnJustin2010} used the emptiness correlation function method to 
 estimate  the arctic curve between the external 
frozen region and an intermediate disordered region that separates it from the bulk 
with AF order. As far as we know, there is no prediction for the internal boundary between the D and AF regions.

\section{Method}
\label{sec:method}

In this section we introduce the key ideas and equations of the belief
propagation (BP) algorithm, which are needed to apply the method to the
present problem. 

We label the lattice edges $i, j, k, \ldots$ and the vertices $a, b, c,
\ldots$ (not to be confused with the vertex weight parameters used above). 
If an edge $i$ is incident on a vertex $a$ we say that the
edge belongs to the vertex and we write $i \in a$. In the language of
probabilistic graphical models (see e.g.\ \cite{CVMreview} and
refs.\ therein for the relationship between these models and
statistical mechanics), edges (respectively, vertices) are variable
nodes (resp.\ factor nodes) in a factor graph. In this notation the
full Boltzmann weight can be written as
\begin{equation}
\psi = \prod_a \psi_a(s_a),
\end{equation}
where $s_a = \{ s_i, i \in a \}$ and $s_i$ represents the orientation
of edge $i$, say, positive towards the right (up) for horizontal (vertical) links, as in~\cite{Syljuasen2004}. 

The BP algorithm is a message--passing algorithm for finding the
minima of an approximate (mean--field--like)
variational free-energy (a generalization of the Bethe--Peierls one)
which can be obtained by truncating the cumulant expansion of the
entropy of the model, by keeping only edge and vertex contributions
or, in other words, the free-energy of the cluster variational method
(see \cite{CVMreview} for a recent review) with vertices as maximal
clusters:
\begin{eqnarray}
{\cal F} &=& \sum_a \sum_{s_a} H_a(s_a) p_a(s_a) + \nonumber \\
&& + k_B T \left[ \sum_a
  \sum_{s_a} p_a(s_a) \ln p_a(s_a) - \sum_i \sum_{s_i} p_i(s_i) \ln
  p_i(s_i) \right],
\label{CVMfree}
\end{eqnarray}
where $H_a(s_a) = -k_B T \ln \psi_a(s_a)$ is the contribution of
vertex $a$ to the model Hamiltonian and $p_a(s_a)$ (respectively
$p_i(s_i)$) is the probability distribution of vertex $a$ (resp.\ edge
$i$). The above variational free-energy has to be minimized with
respect to the vertex and edge probability distributions, subject to
the constraints of normalization
\begin{equation}
\sum_{s_a} p_a(s_a) = 1, \quad \forall a \qquad \sum_{s_i} p_i(s_i) = 1,
\quad \forall i
\end{equation}
and marginalization
\begin{equation}
\sum_{s_{a \setminus i}} p_a(s_a) =  p_i(s_i), \quad \forall a, \quad \forall i
\in a,
\label{CVMmarg}
\end{equation}
where $s_{a \setminus i} = \{ s_j, j \in a, j \ne i \}$.

Minima of the variational free-energy correspond
\cite{Yedidia2003,CVMreview} to fixed points of the BP
algorithm. In the latter picture, a vertex $a$ sends a message $m_{a
  \to i} (s_i)$ to an edge $i \in a$, which depends on the edge
orientation. Up to a normalization constant, the edge and vertex probability
distributions can be written as functions of the messages as
\begin{eqnarray}
p_i(s_i) \propto \prod_{a \ni i} m_{a \to i}(s_i), \qquad\qquad
p_a(s_a) \propto \psi_a(s_a) \prod_{i \in a} \prod_{b \ni i}^{b \ne a} 
m_{b \to i}(s_i). 
\label{CVMbeliefs}
\end{eqnarray}
Rewriting the marginalization condition Eq.\ (\ref{CVMmarg}) in terms of
messages one obtains, again up to a normalization constant,
\begin{equation}
m_{a \to i}(s_i) \propto \sum_{s_{a \setminus i}} \psi_a(s_a) \prod_{j \in a}^{j \ne i} \prod_{b \ni j}^{b \ne a} 
m_{b \to j}(s_j),
\end{equation}
the main iterative equation of the BP algorithm, the fixed point of which
  gives the message values at equilibrium, to be used in the
  computation of the edge and vertex probability distributions,
  Eq.\ (\ref{CVMbeliefs}), and of the free-energy,
  Eq.\ (\ref{CVMfree}). In the following we shall often refer to the
  free-energy density $f = {\cal F}/(NM)$.
 
Domain--wall boundary conditions are easily represented in this
scheme, by introducing boundary messages, sent by auxiliary vertices
to boundary edges which vanish if and only if the edge orientation
differs from the one imposed by the boundary conditions. In more detail, what we mean by this is the following.
Consider a boundary edge $i$ and assume that  it is horizontal, on the right boundary of the lattice. This edge belongs to
  a single vertex $a$, which sits on its left side. Introduce the auxiliary vertex 
  $b$, sitting on the right of the edge $i$. Since the polarization of $i$ is constrained to be
  rightward ($+1$ in our representation), we set $m_{b \to i}(-1) = 0$.

\section{Results}
\label{sec:results}

In this Section we present the results that we obtained by using the
BP algorithm explained in Sec.~\ref{sec:method}.  

\subsection{Square lattice}

We first study the model 
on a square lattice for which analytic expressions for the interfaces are 
known.

\subsubsection{The free-energy.}

In~\cite{Foini2013} the Bethe-Peierls approximation was used to study 
the equilibrium properties of the sixteen and, in particular, the six vertex model with periodic
boundary conditions. Defining the six vertex model on a tree,  the exact FE free-energy 
density was found, while in the  D and AF phases very good approximations 
to the exact expressions, that get better and better far from the transition, were derived. 
Although the free-energy densities obtained with the Bethe-Peierls approximation in the infinite size limit 
(that in the homogeneous case can be treated analytically) are
not exact, the location of the different transition lines (between D and FE, and D and AF,
phases) are. 

\begin{figure}[h]
\centerline{\includegraphics[width=0.7\textwidth]{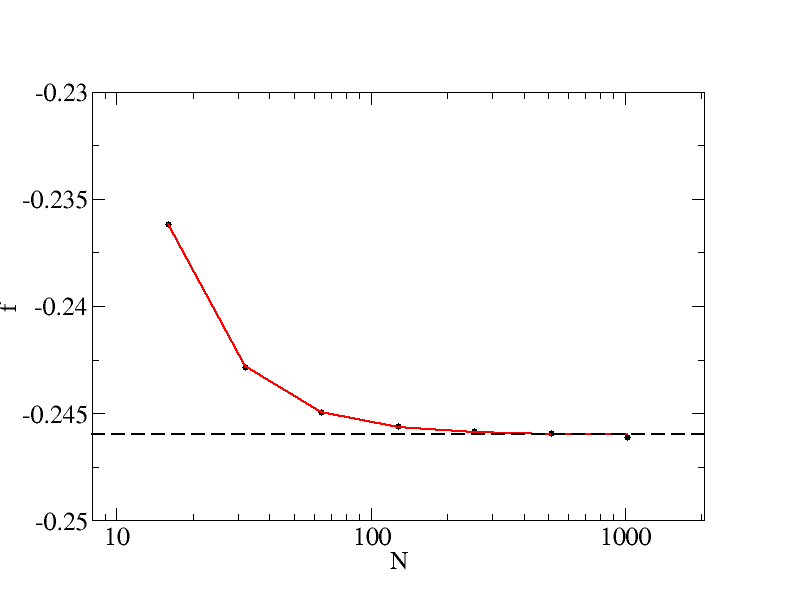}}
\caption{BP free-energy density vs.\ lattice size $N$ for the six
  vertex model on a square lattice with domain-wall boundary
  conditions at the spin--ice point, $a=b=c =1$ (data
  points joined by straight lines). The asymptotic estimate, $f_{DW}^D
  \simeq -0.246$, is shown with a dotted horizontal lines.  The exact
  value, $f^D_{DW}\simeq -0.262$, lies beyond the vertical range of
  variation of this graph.  }
\label{fig:fSpinIce}
\end{figure}

First, we check that the BP free-energy densities of the six-vertex 
model with DWBC satisfies $f_{DW}>f_{P}$ for a square lattice system. In Fig.~\ref{fig:fSpinIce} we show the approach to the
thermodynamic limit of the free-energy density at the spin-ice point
$a=b=c=1$ ($\Delta=1/2$). The numerical data are fitted with the form
\begin{equation}
f_{DW}^D(N) \simeq f_{DW}^D(N\to\infty) + c N^{-\psi} 
\; .
\end{equation}
We find $\psi\simeq 1.6 - 1.7$ and $f_{DW}^D(N\to\infty) \simeq
-0.246$. The estimate for the free-energy density in the thermodynamic
limit is larger than the one for periodic boundary conditions,
$f_{P}^D = -\ln(3/2)\simeq -0.404$, that is Pauling's result also
found on the tree.  We note that the free-energy-density for
domain-wall boundary conditions on the tree is slightly larger
than the exact result, $f_{DW}^D = - \ln
(3\sqrt{3}/4) \simeq -0.262$~\cite{KorepinZinnJustin00}. The fact that
the BP approximation slightly over-estimates the free-energy (with
respect to the exact result) was also found for periodic boundary
conditions~\cite{Foini2013}.

\begin{figure}
\centerline{\includegraphics[width=0.7\textwidth]{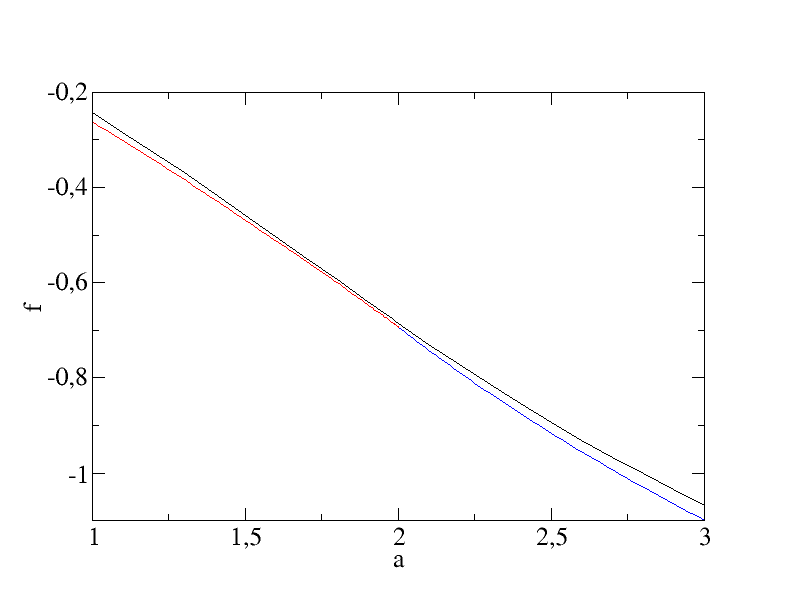}}
\caption{Free-energy density vs. $a$ for the six vertex model on a square lattice with 
parameters $b = c = 1$ and under domain-wall boundary conditions. The BP data for a 
system with linear size $N = 32$
  (black line) should be compared to the exact result~\cite{KorepinZinnJustin00} in the thermodynamical
  limit (red line: D phase, blue line: FE phase).}
\label{fig:DvsFE}
\end{figure}

In Fig.\ \ref{fig:DvsFE} we compare our free-energy density with the
exact one \cite{KorepinZinnJustin00}, in the D and FE phases. In
particular we consider a square lattice with $N=32$, we set $b = c =
1$ and vary $a$ in the range $[1,3]$. Notice that $a=1$ corresponds to
the spin--ice point $\Delta = 1/2$, while at $a=2$ ($\Delta = 1$) the
system is expected to undergo (in the thermodynamical limit) a
continuous phase transition between the D and FE phases. Our finite
size analysis shows instead a continuous variation from a D bulk
behaviour for $a < 2$ to an FE bulk behaviour for $a > 2$. In
agreement with data reported in Fig.\ \ref{fig:fSpinIce}, the
corrections to our free-energy density that we obtain upon increasing
$N$ from 32 to 1000 are of order $10^{-3}$, hence not resolvable on the scale of
Fig.\ \ref{fig:DvsFE}.

The exact FE free-energy of the two-dimensional model in the $N\to\infty$
  limit is $f^{FE}_{DW}=f^{FE}_P= -\ln a$ and it is shown with a blue line in Fig.~\ref{fig:DvsFE}. 
 The BP algorithm should also give this result in the infinite size limit.
 We ascribe the deviation from of the black curve in Fig.~\ref{fig:DvsFE} from the
 asymptotic limit to finite size effects.
  
\begin{figure}
\centerline{\includegraphics[width=0.7\textwidth]{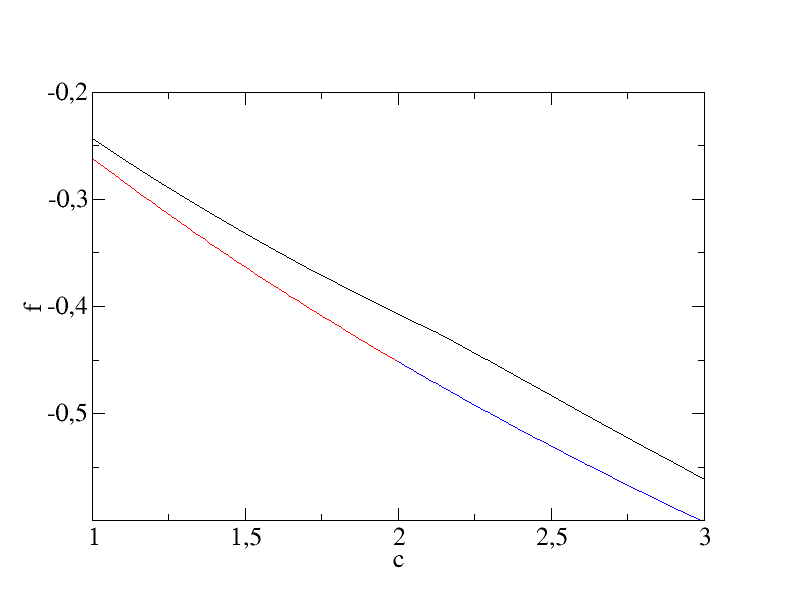}}
\caption{Free-energy density vs. $c$ for the six vertex model on a square lattice with 
parameters $a  = b = 1$ and under domain-wall boundary conditions. The BP data for a 
system with linear size $N = 32$
  (black line) should be compared to the exact result~\cite{KorepinZinnJustin00,ZinnJustin2000} in the thermodynamical
  limit (red line: D phase, blue line: AF phase).}
\label{fig:DvsAF}
\end{figure}

A similar analysis is shown in Fig.\ \ref{fig:DvsAF} in the D and AF
phases (for the exact free-energy density in the AF phase see
\cite{ZinnJustin2000}). We set $N = 32$, $a = b = 1$ and vary $c$ in
the range $[1,3]$. Again $c = 1$ corresponds to the spin--ice point
$\Delta = 1/2$, while at $c=2$ ($\Delta = -1$) the system is expected
to undergo (in the thermodynamical limit) a continuous phase
transition between the D and AF phases. Here our finite size results
show a reminiscence of the infinite size phase transition. For $c \in
[2.08,2.13]$ we find both a (stable) AF phase and a (metastable) D
phase (the difference in free-energy density between these two phases is
$\sim 10^{-4}$, not resolvable on the scale of
Fig.\ \ref{fig:DvsAF}), while for $c < 2.08$ we have only the D phase and for $c >
2.13$ only the AF phase is found. This allows us to locate a finite
size phase transition at $c \simeq 2.08$. Upon increasing $N$ the
transition point seems to tend towards $c = 2$, but it is hard to
prove that the transition line is captured exactly as $N \to \infty$.

\subsubsection{Arctic circle.}

\begin{figure}[h]
\centerline{\includegraphics[width=0.7\textwidth]{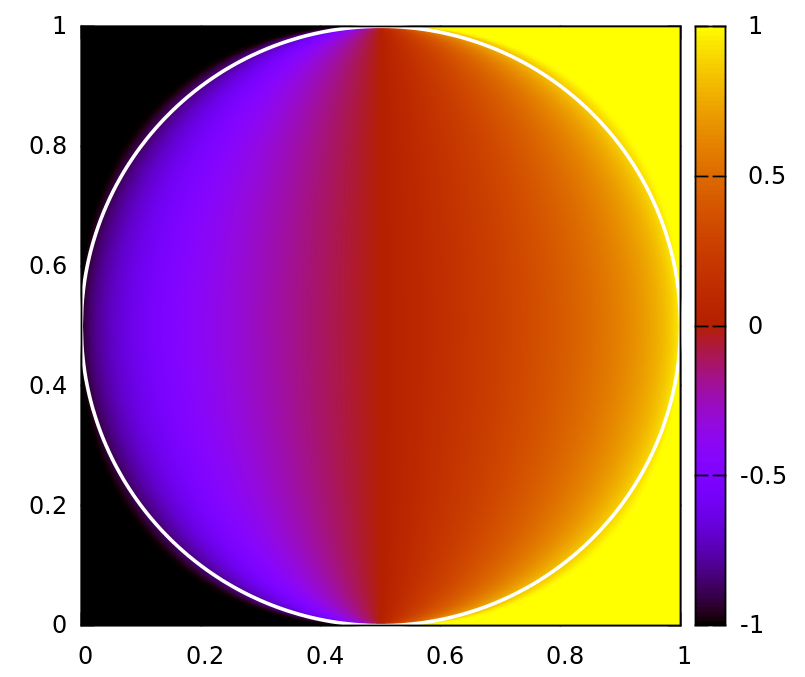}}
\caption{BP polarization of the horizontal edges for the free-fermion case, $\Delta = 0$, with $a
  = b$, on a square lattice with $1024 \times 1024$ lattice vertices. The white line is the exact
  arctic circle. }
\label{fig:Circle}
\end{figure}

\begin{figure}[h]
\centerline{\includegraphics[width=0.7\textwidth]{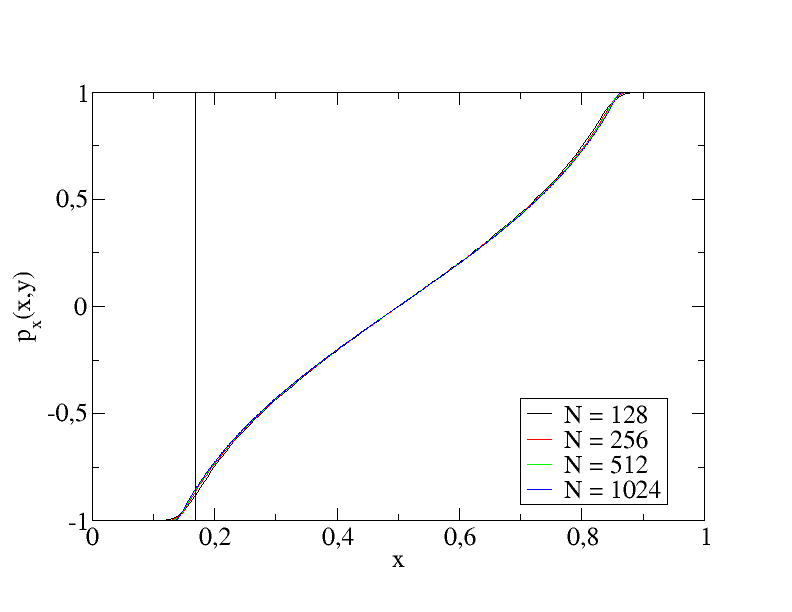}}
\caption{BP polarization of the horizontal edges for the free-fermion
  case, $\Delta = 0$, with $a = b$, for $y = 1/8$ and different system
  sizes $N$. The vertical line denotes the left intersection of the
  exact arctic circle with $y = 1/8$.}
\label{fig:CircleSection}
\end{figure}

We work out the case $\Delta=0$ and $a=b$ in which the arctic curve
should be a circle. In Fig.\ \ref{fig:Circle} we plot the polarization
$\langle s_i \rangle$ of the horizontal edges for a $1024 \times 1024$
lattice, with (following the convention introduced in
Sec.\ \ref{sec:method} above) $s_i = +1$ (respectively, $-1$) for a
rightward (resp.\ leftward) edge. The exactly known arctic circle is
drawn as a white solid line. The agreement is remarkable, considering
that we are applying a mean--field like technique to a system in its
critical state.  The temperate region is slightly overestimated by the
BP approximation, though this might be just a finite size effect. This
is better observed in Fig.\ \ref{fig:CircleSection}, where we plot the
polarization (from now on, the polarization of the horizontal edge at
position $(x,y)$ will be denoted by $p_x(x,y)$) of the horizontal
edges at a fixed vertical coordinate $y = 1/8$, for different lattice
sizes, together with the left intersection of the exact arctic circle
with $y = 1/8$.

\begin{figure}[h]
\centerline{
\includegraphics[width=0.53\textwidth]{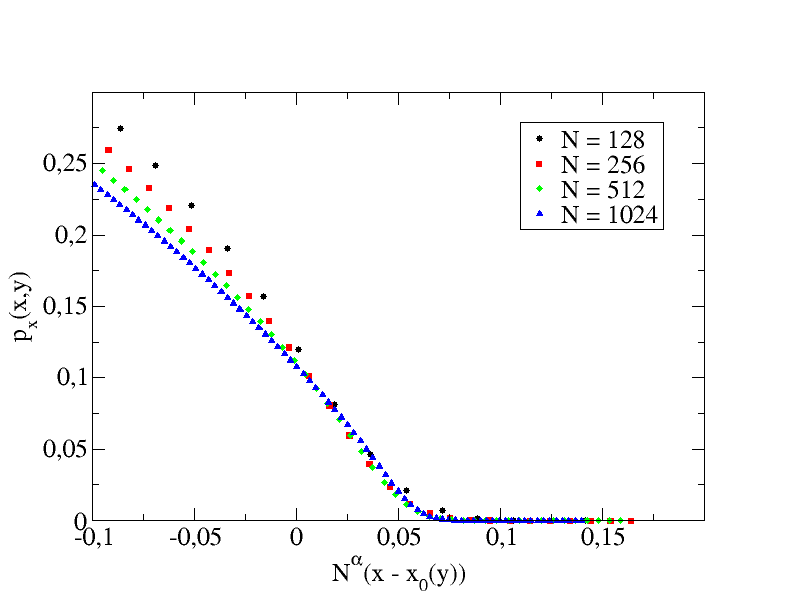}
\includegraphics[width=0.53\textwidth]{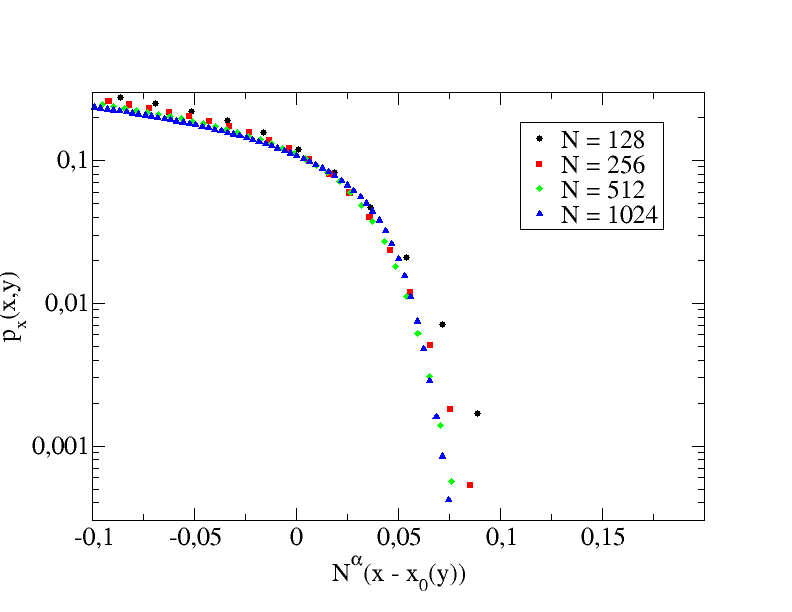}
}
\caption{BP polarization (deviation from the value in the frozen
  phase) of the horizontal edges vs.\ $N^\alpha (x - x_0(y))$ for the
  free-fermion case, $\Delta = 0$, with $a = b$, for $y = 3/4$,
  $\alpha = 1/6$ and different system sizes $N$. Vertical axis in
  linear (left panel) and log (right panel) scale.}
\label{fig:CircleScaling}
\end{figure}

Given that the system is critical, we expect simple power--laws for
the scaling of its observables. We denote by $x_0(y)$ the (exactly
known) arctic curve. Observing that in the frozen regions on the right
of Fig.\ \ref{fig:Circle} the polarization $p_x(x,y)$ tends to 1, we
have considered the deviation $1 - p_x(x,y)$ as a function of
$N^\alpha (x - x_0(y))$ for fixed $y$ and $x > 1/2$ and looked for
data collapse. The best results have been obtained with $\alpha \simeq
1/6$ and the corresponding plot is reported in
Fig.\ \ref{fig:CircleScaling}, with the vertical axis both in linear
(left panel) and log (right panel) scale. A good collapse is obtained
for $x \gtrsim x_0(y)$ suggesting that our estimate of the arctic
curve is very close to the exact one and that the scaling law holds in
the frozen phase only.

\subsubsection{Arctic ellipses.}

\begin{figure}[h]
\centerline{\includegraphics[width=0.7\textwidth]{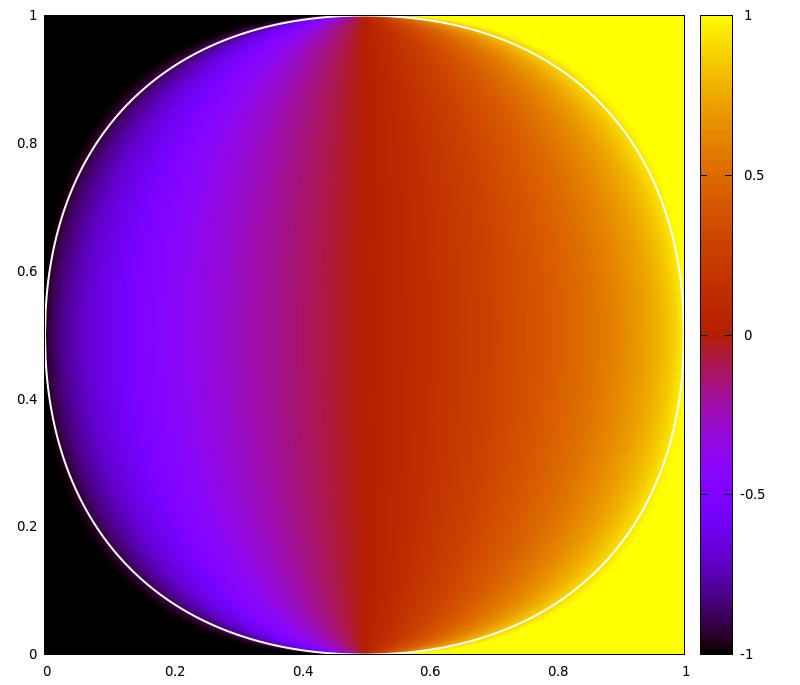}}
\caption{BP polarization of the horizontal edges at the spin--ice point, $a = b = c$,
  $\Delta = 1/2$.}
\label{fig:Ellipse}
\end{figure}

\begin{figure}[h]
\centerline{
\includegraphics[width=0.52\textwidth]{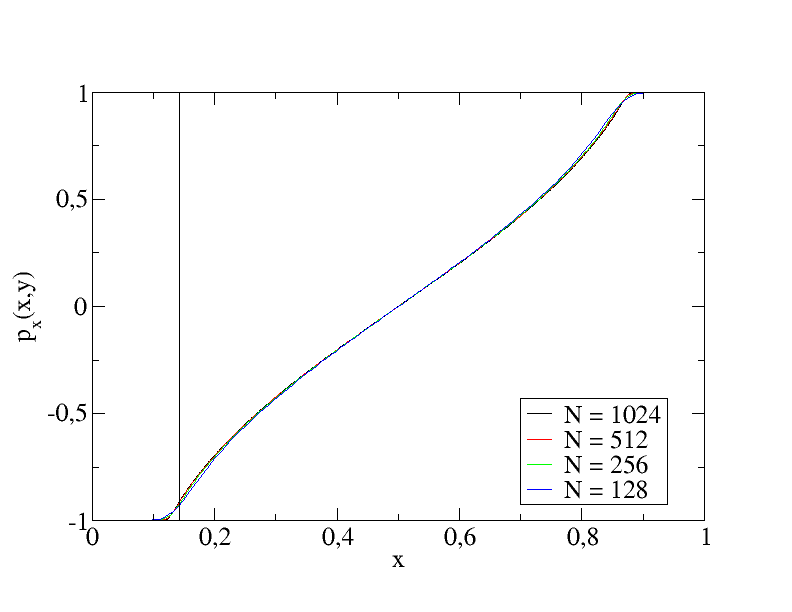}
\includegraphics[width=0.52\textwidth]{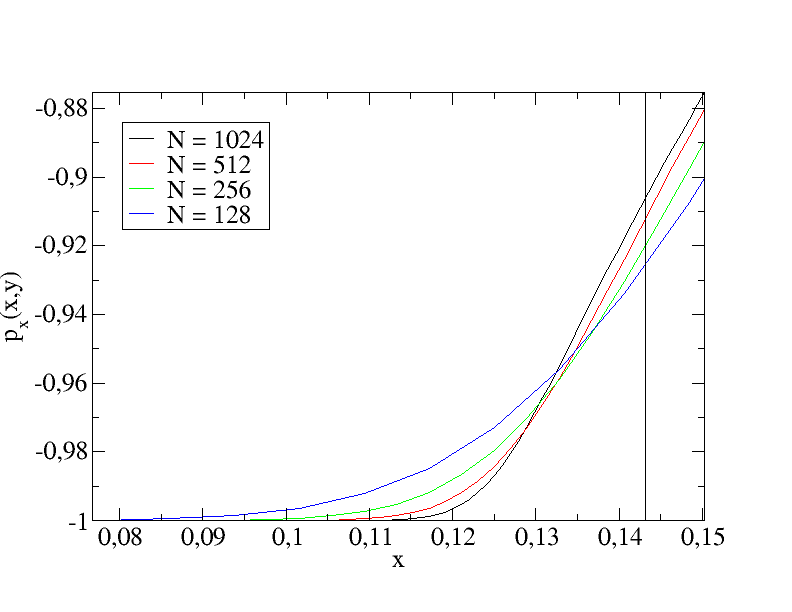}
}
\caption{Left panel: BP polarization of the horizontal edges at
    the spin--ice point, $a = b = c$, $\Delta = 1/2$, for $y = 1/8$
    and different system sizes $N$. The vertical line denotes the left
    intersection of the exact arctic ellipse with $y = 1/8$. Right
    panel: magnification close to the arctic ellipse. 
    }
\label{fig:Sections}
\end{figure}

We now work out the case $\Delta=1/2$ and $a=b=c$ in which the arctic
curve should be an ellipse. In Fig.\ \ref{fig:Ellipse} we plot, as in
the previous subsection, the polarization of the horizontal edges and
the exactly known arctic ellipse.  The agreement is still remarkable.
Our numerical data are consistent with the exact $\kappa=1/2$ values for the 
contact points. However, also in this case the temperate region seems slightly overestimated
by the BP approximation. Indeed, in Fig.\ \ref{fig:Sections}. we display the polarisation 
of the horizontal edges as a function of the horizontal coordinate for a cut at constant 
vertical coordinate, $y=1/8$. The right panel shows a zoom close to the $x$ value where the 
exact article curve lies. One sees that the finite $N$ curves obtained with the BP algorithm 
suggest a location of the sharp boundary at a value of $x$ that will be closer to the 
edge of the system than the exact one.

\subsubsection{Double interfaces in the AF phase.}

\begin{figure}[h]
\centerline{
\includegraphics[width=0.33\textwidth]{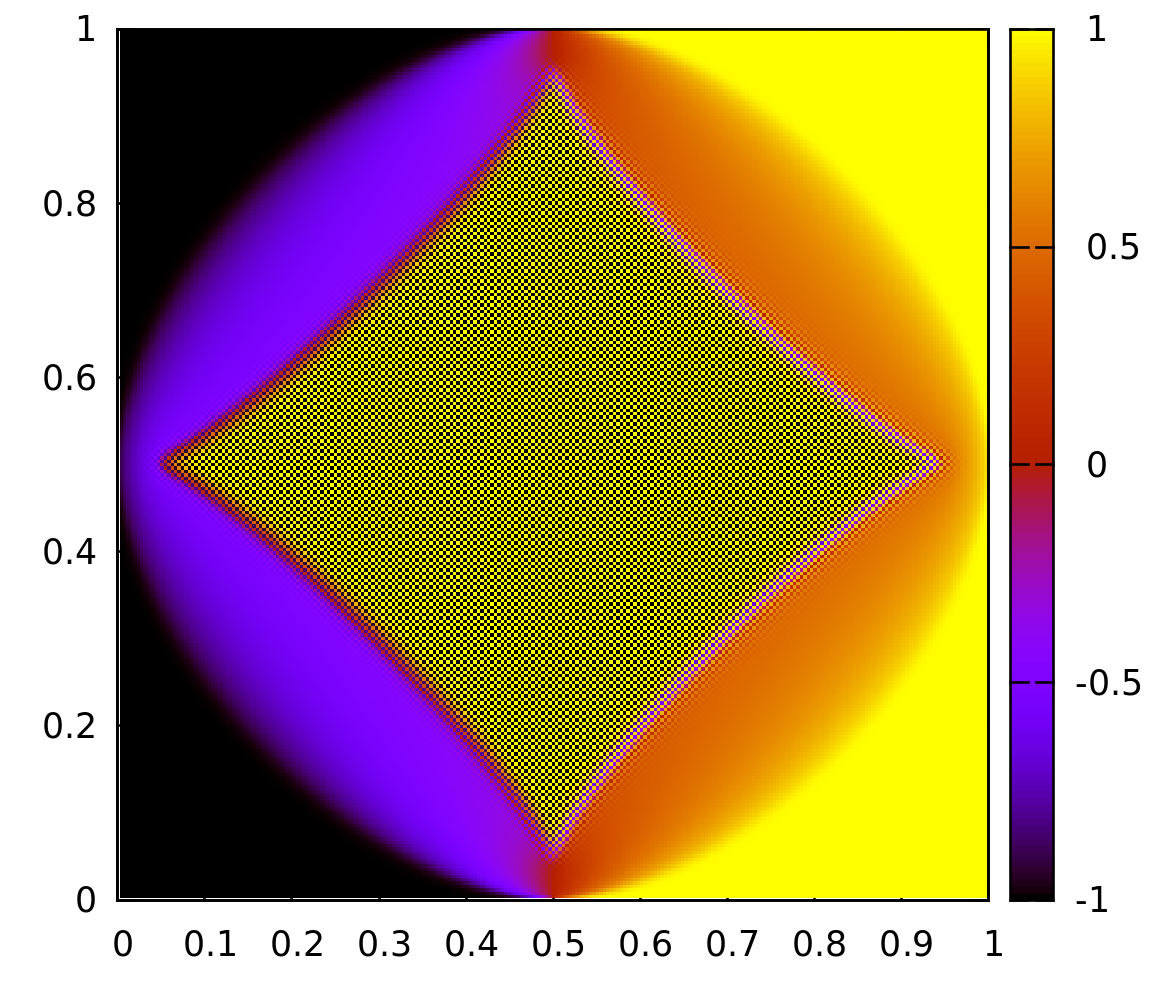}
\includegraphics[width=0.33\textwidth]{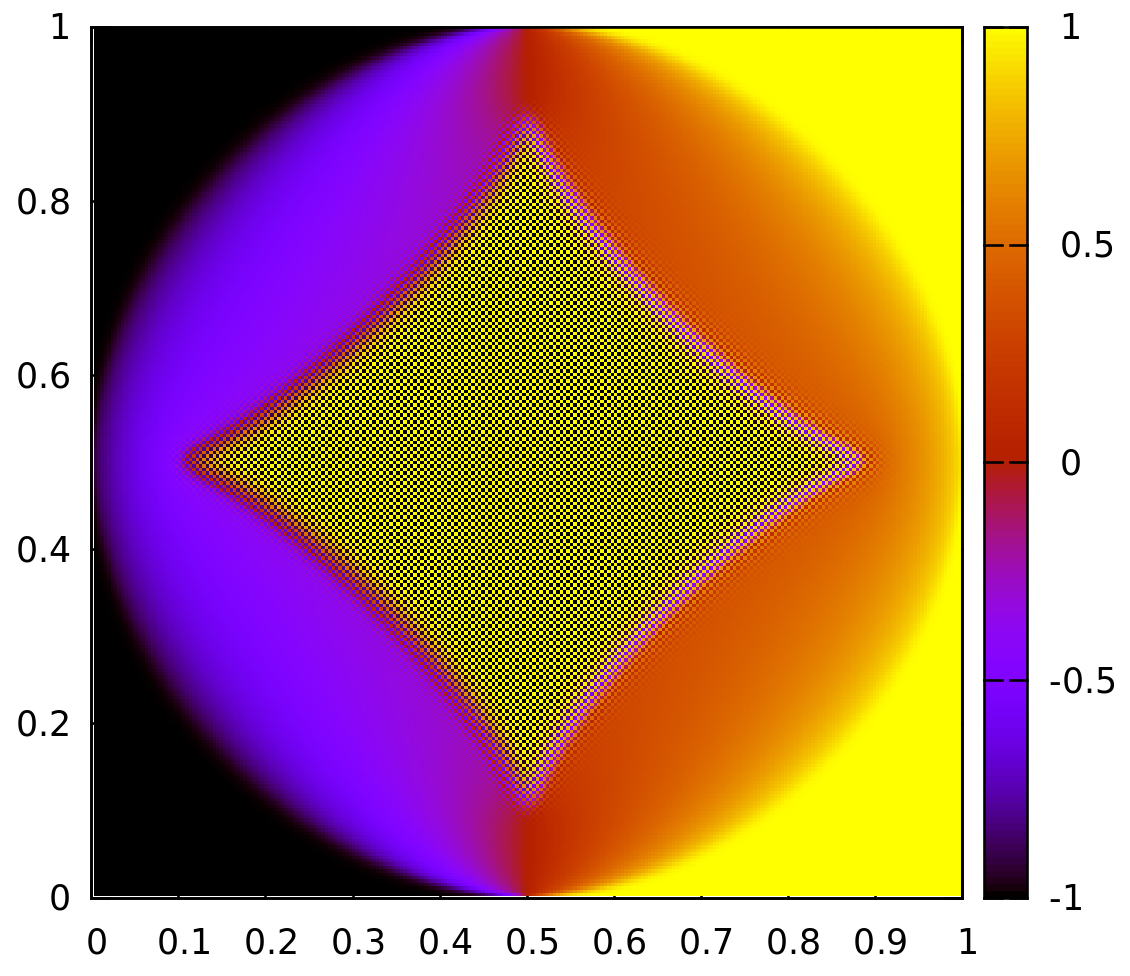}
\includegraphics[width=0.33\textwidth]{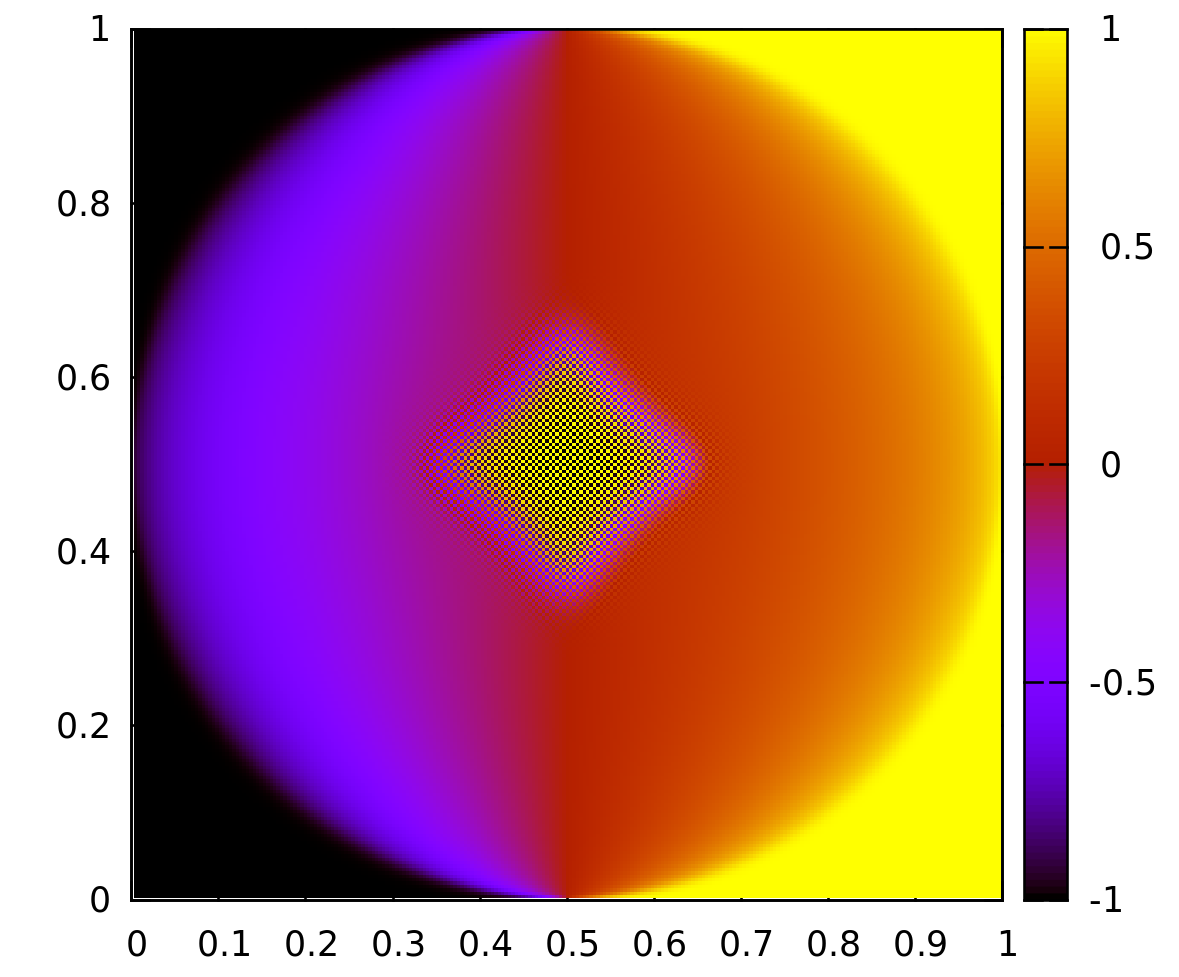}
}
\caption{BP polarization of the horizontal edges for
    $\Delta <-1$ with $a=b=1$ and $c=3, 2.5, 2.05$ from left to
    right. The system size is $N = 256$.}
\label{fig:AF}
\end{figure}

For bulk parameters such that the system is in the AF phase, the
domain wall boundary conditions impose a frozen external region, an
intermediate disordered `buffer' and an internal region with AF
order. This is shown in Fig.~\ref{fig:AF} for different values of $c >
2$ and $a=b=1$ (the phase transition between the bulk D and AF phases is
located at $c = 2$).

The shrinking of the internal AF region as $c$ decreases towards 2
suggests a continuous transition: indeed, this transition is known to
be continuous, and in particular of infinite order
\cite{ZinnJustin2000} (a feature which of course cannot be reproduced
by our mean--field--like approach).

\begin{figure}[h]
\centerline{
  \includegraphics[width=1\textwidth]{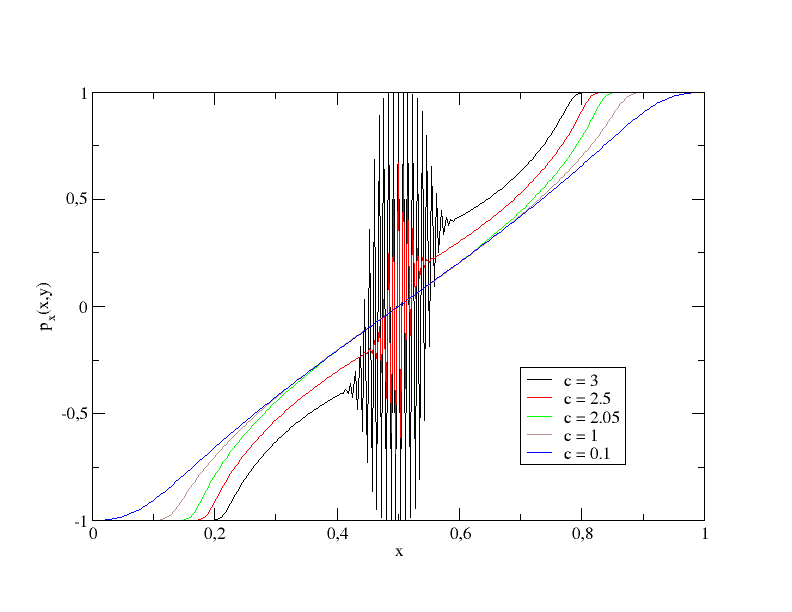}}
\caption{BP polarization of the horizontal edges at $a = b = 1$ meaning
$\Delta = (2-c^2)/2$, system size $N = 256$, and $y = 1/8$, for
different $c$ values.  For $c=3, \ 2.5, \ 2.05$, $\Delta < -1$ and
  the system prefers the AF configuration in the bulk. For $c=1,
  \ 0.1$, $0< \Delta <1 $ and the system prefers the D configuration
  in the bulk.}
\label{fig:Trend}
\end{figure}

While the variation of the AF--D boundary with $c$ is evident from
Fig.\ \ref{fig:AF}, it is not the same for the D--frozen
boundary (the arctic curve). We can however understand its trend from
Fig.\ \ref{fig:Trend}, where we have plotted the polarization of the
  horizontal edges at a fixed height $y = 1/8$ for various values of
  $c$. Looking at the points where the polarization saturates it is
  now evident that the arctic curve gets larger as $c$ decreases, both
  for $c > 2$ (bulk AF phase, curves characterized by oscillations in
  the center) and for $c < 2$ (bulk D phase, smooth curves). 

\subsection{Rectangular systems with partial domain wall boundary conditions}

In the framework of the BP algorithm it is straightforward to consider
rectangular lattices and different types of boundary conditions. In
order to illustrate this, in the present Section we discuss the
six vertex model on rectangular lattices, with partial domain wall
boundary conditions (pDWBC). These are a generalization of DWBC to
rectangular systems \cite{Foda2012,Bleher2014}. Consider a rectangular
lattice with $N \times M$ vertices (in the following we shall assume
$N > M$ to fix ideas) and impose DWBC--like boundary conditions on
three sides of the lattice, leaving the $4^{\rm th}$ side
free. Following the convention we have set up in Fig.\ \ref{fig:dwbc}
we will have inward--pointing edges on the bottom side and
outward--pointing edges on the left and right sides. The ice rule then
implies that on the top side we must have $M$ inward--pointing edges
and $N-M$ outward--pointing edges (hence for $M = N$ DWBC are
recovered). The condition that the top side is left free is equivalent
to sum over all possible distributions of its $M$ inward--pointing
edges.

\begin{figure}[h]
\centerline{
\includegraphics[width=0.33\textwidth]{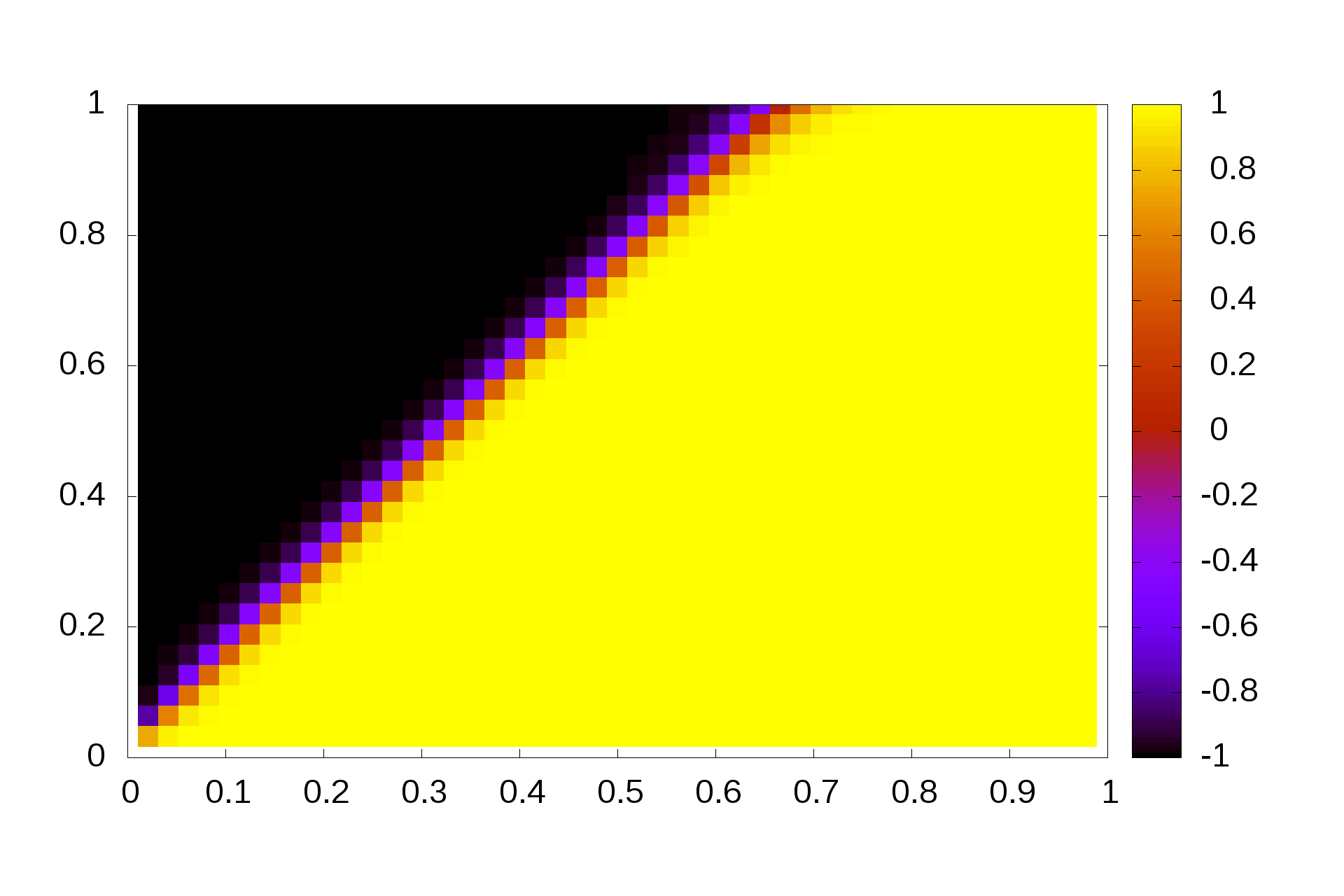}
\includegraphics[width=0.33\textwidth]{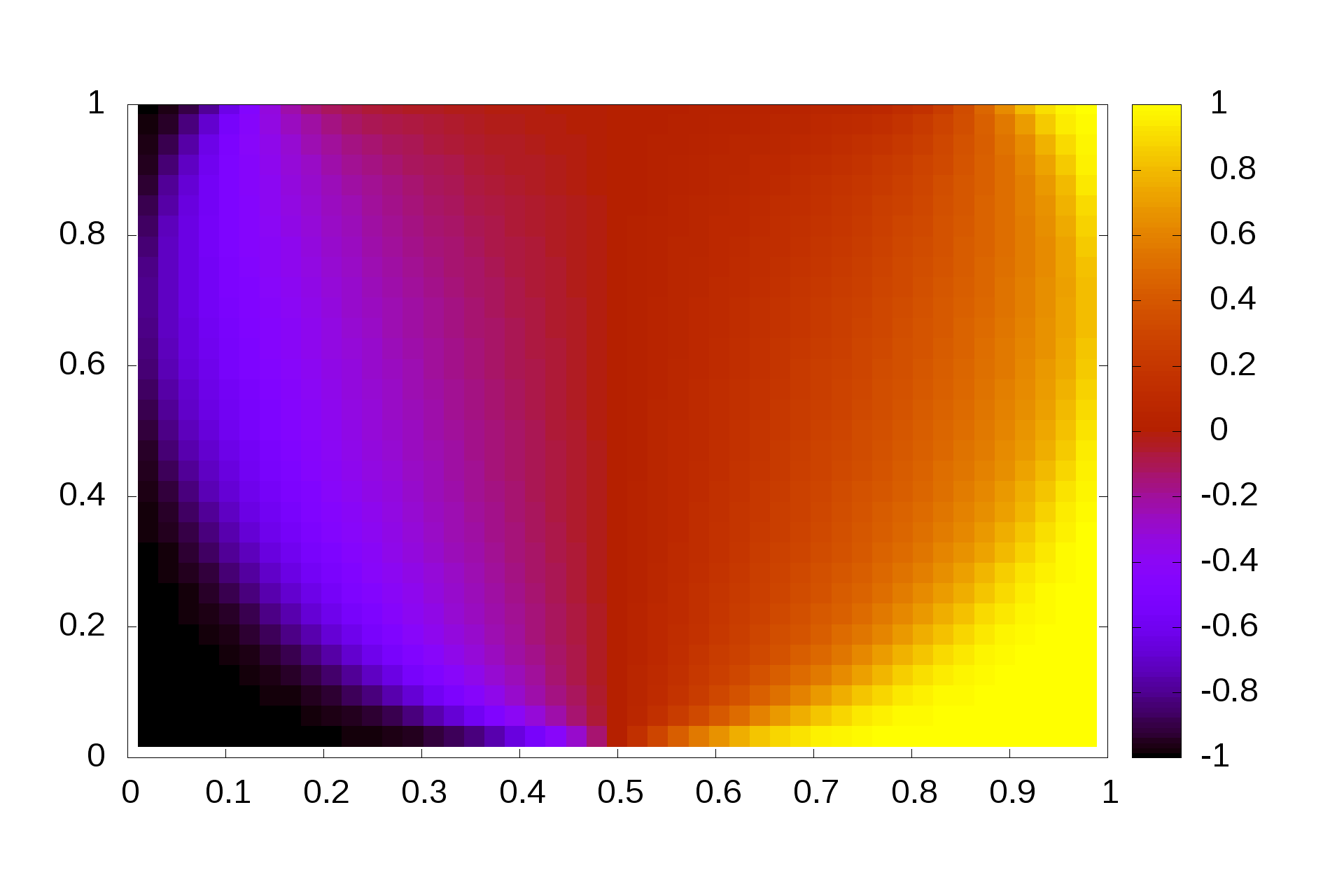}
\includegraphics[width=0.33\textwidth]{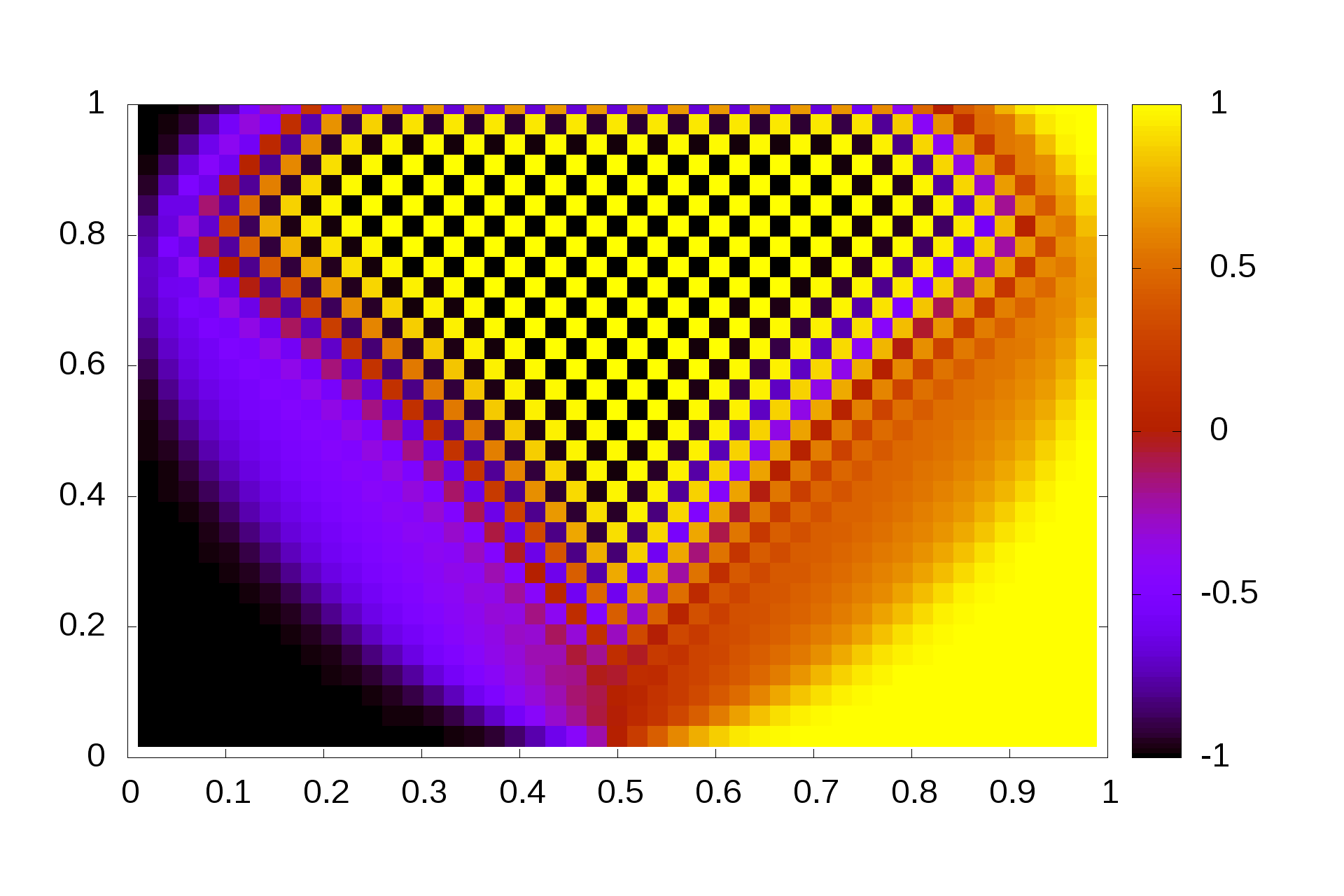}
}
\caption{BP polarization of the horizontal edges for a $48 \times 32$
  rectangular system and $(a,b,c) = (3,1,1)$ (FE phase, left panel),
  (1,1,1) (D phase, center panel) and (1,1,2.5) (AF phase, right
  panel).}
\label{fig:pDWBC}
\end{figure}

The effect of pDWBC is illustrated in Fig.\ \ref{fig:pDWBC} with a $48
\times 32$ system and different values of the parameters,
corresponding to the FE, D and AF phases from left to right.
 
In the FE phase the effect of pDWBC is trivial: the interface between
the two symmetric homogeneous phases remains at $45^\circ$ with
respect to the lattice boundary, and the $M$ inward--pointing edges of
the top side are, in the most probable arrangement, placed on the
right. In the D and AF phase the phase separation phenomena and the
arctic line are still observed, although with a different geometry
which, qualitatively, can be thought of as obtained by cutting the top
16 rows of a square $48 \times 48$ lattice.

\begin{figure}[h]
\centerline{\includegraphics[width=0.5\textwidth]{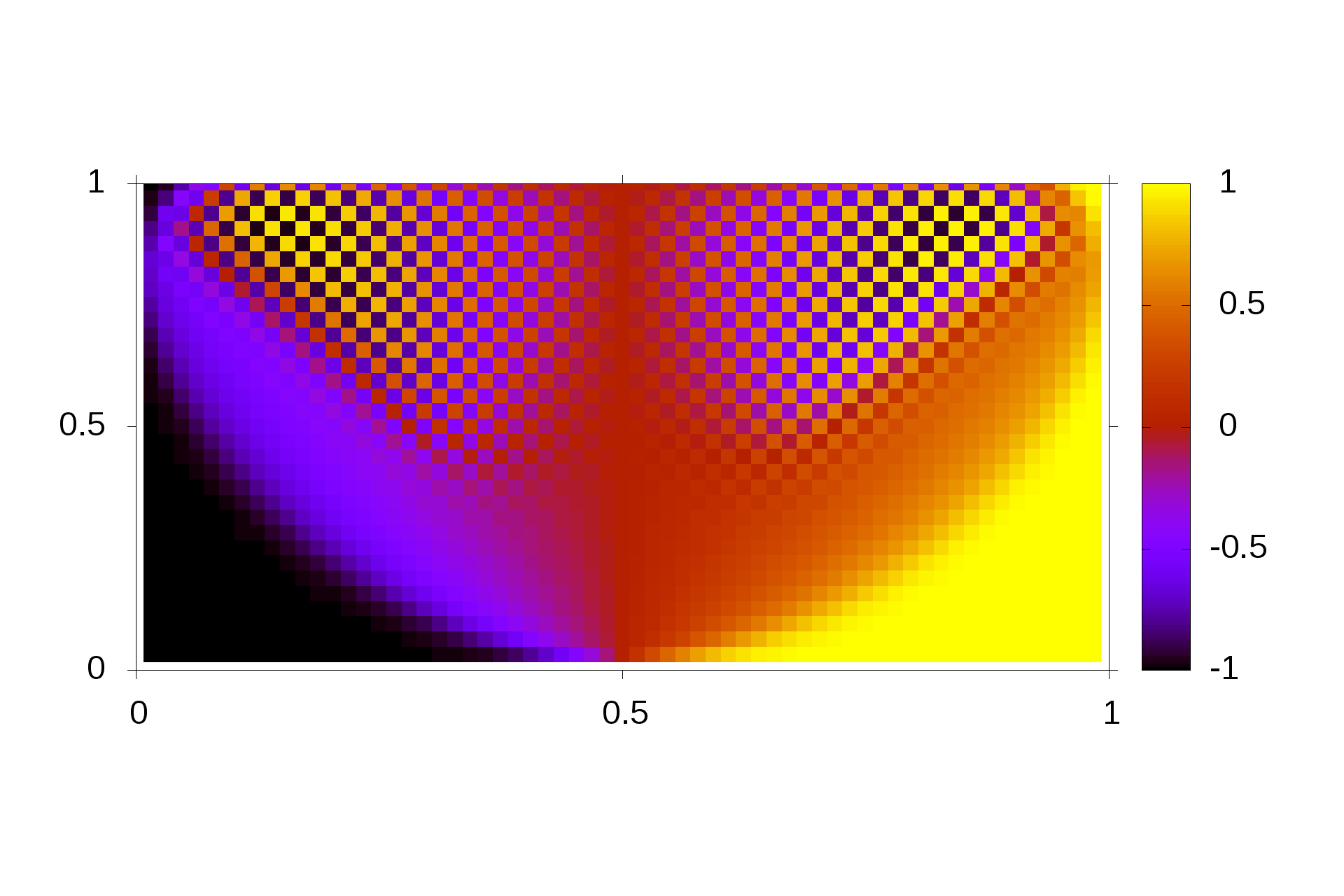}}
\centerline{\includegraphics[width=0.75\textwidth]{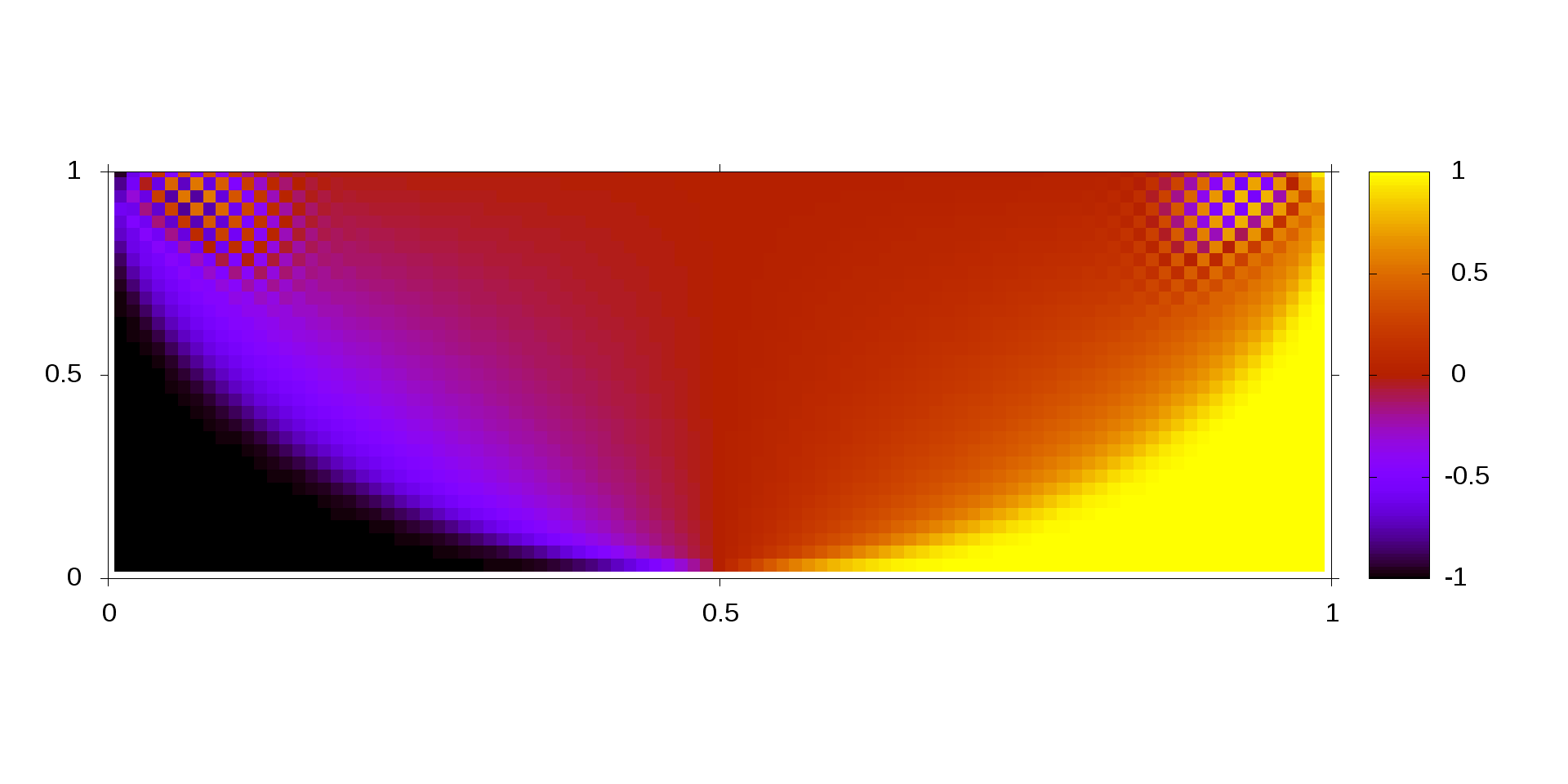}}
\centerline{\includegraphics[width=\textwidth]{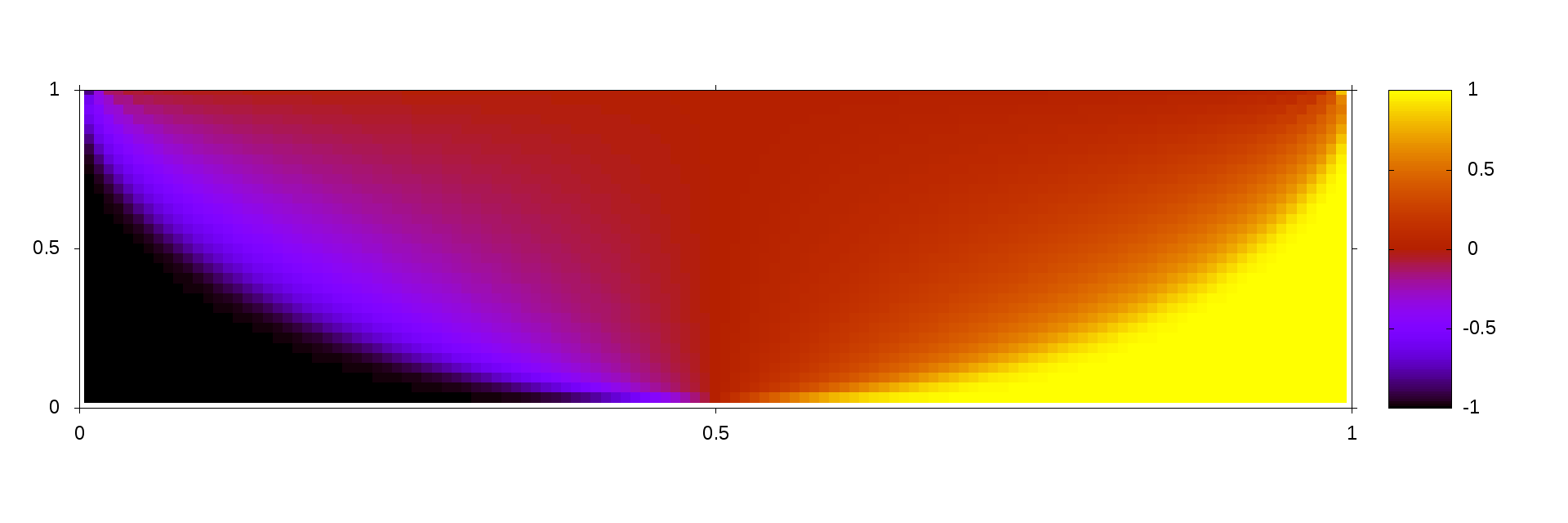}}
\caption{BP polarization of the horizontal edges in the AF phase for
  $a = b = 1$, $c = 2.5$ for rectangular systems of size $64 \times
  32$ (left panel), $96 \times 32$ (center panel) and $128 \times 32$
  (right panel).}
\label{fig:AspectRatioAF}
\end{figure}

The most interesting phenomena are observed by varying the aspect
ratio in the AF phase, as shown in Fig.\ \ref{fig:AspectRatioAF}.

As the aspect ratio decreases the inner AF region is first split in 2
regions, which then get farther from each other and eventually
disappear. 

\section{Outlook}
\label{sec:conclusions}

We studied the spatial organisation of vertices in the six vertex
model with domain wall boundary conditions by using an approximate
method, the belief propagation or Bethe-Peierls technique. We found
that, although the method is not expected to give exact results for the
location of the various arctic and temperate curves, the forms obtained are
remarkably close to the analytic ones, when these are known. The
advantage of this method is that it is quite simple to implement and
that it can be applied to situations in which the various exact
methods used in the literature do not necessarily apply. For instance,
any other type of fixed boundary conditions \cite{Zinn-Justin2002}, or
asymmetric weights ($\omega_1 \ne \omega_2$, $\omega_3 \ne \omega_4$, $\omega_5 \ne \omega_6$) could be
easily dealt with using the same technique. 

Our study has been static. It would be interesting to analyse the dynamics of 
formation of such structures allowing for the presence of a small density of 
defects, {\it e.g.} along the lines in~\cite{Budrikis10,Levis2012,Budrikis12,Levis2013b}, and 
with protocols as the ones used experimentally~\cite{Nisoli10,Morgan10,Morgan13}. 
The analysis of finite size systems with various aspect ratios will be especially 
relevant to compare with experiments.

Artificial spin-ice samples can be imaged in real space with various magnetic microscopy such as Lorentz 
microscopy~\cite{PhysRevB.83.174431}, 
photoelectron emission microscopy~\cite{PhysRevB.78.144402,PhysRevLett.111.057204} and 
magnetic force microscopy~\cite{Ladak10}.  With these techniques one could access the spatial structures that we discussed in this 
work.

\appendix

\vspace{0.25cm}
\noindent
{\bf Acknowledgements.}
We thank C. H. Marrows for exchanges on the possible experimental realisation of domain wall boundary conditions in 
artificial two dimensional spin-ice samples. We warmly thank  F. Colomo, A. Sportiello and P. Zinn-Justin for very useful discussions.
 L.~F.~C. is a member of Institut Universitaire de France.

\section*{References}
\bibliographystyle{iopart-num}
\bibliography{MC-Bethe}

\providecommand{\newblock}{}
\begin{thebibliography}{10}
\expandafter\ifx\csname url\endcsname\relax
  \def\url#1{{\tt #1}}\fi
\expandafter\ifx\csname urlprefix\endcsname\relax\def\urlprefix{URL }\fi
\providecommand{\eprint}[2][]{\url{#2}}

\bibitem{Elkies1992a}
Elkies N, Kuperberg G, Larsen M and Propp J 1992 {\em J. Algebraic Combin.|\/}
  {\bf 1} 111

\bibitem{Elkies1992b}
Elkies N, Kuperberg G, Larsen M and Propp J 1992 {\em J. Algebraic Combin.|\/}
  {\bf 1} 219

\bibitem{Cohn1996}
Cohn H, Elkies N and Propp J 1996 {\em Duke Math. J.\/} {\bf 85} 117

\bibitem{Jockusch1998}
Jockusch W, Propp J and Shor P Random domino tilings and the arctic circle
  theorem arXiv:math.CO/9801088

\bibitem{Cohn1998}
Cohn H, Marsen L and Propp J 1998 {\em New York J. Math\/} {\bf 4} 137

\bibitem{Borodin2010}
Borodin A, Gorin V and Rains E~M 2010 {\em Selecta Math. (N.S.)\/} {\bf 16} 731

\bibitem{Kenyon}
Kenyon R 2008 Lectures on dimers {\em Exact methods in low dimensional
  statistical physics and quantum computing\/} Les Houches Summer School 89 ed
  Jacobsen J, Ouvry S, Pasquier V, Serban D and Cugliandolo L~F (Oxford
  University Press)

\bibitem{ColomoPronko-proc}
Colomo F and Pronko A~G 2008 The arctic curve revisited {\em Conference on
  Integrable Systems, Random Matrices, and Applications\/} ({\em Contemporary
  Mathematics\/} vol 458) ed Baik J, Kriecherbauer T, Li L and et~al p 361

\bibitem{ColomoPronko2010c}
Colomo F and Pronko A~G 2010 {\em SIAM J. Discrete Mathematics\/} {\bf 24} 1558

\bibitem{ColomoPronko2010b}
Colomo F and Pronko A~G 2010 {\em J. Stat. Phys.\/} {\bf 138} 662

\bibitem{ColomoPronkoZinnJustin2010}
Colomo F, Pronko A~G and Zinn-Justin P 2010 {\em J. Stat. Mech.\/}  L03002

\bibitem{ColomoNoferiniPronko}
Colomo F, Noferini V and Pronko A~G 2011 {\em J. Phys. A: Math. Theor.\/} {\bf
  44} 195201

\bibitem{Lieb1967a}
Lieb E~H 1967 {\em Phys. Rev. Lett.\/} {\bf 18} 692

\bibitem{Lieb1967b}
Lieb E~H 1967 {\em Phys. Rev. Lett.\/} {\bf 18} 1046

\bibitem{Lieb1967c}
Lieb E~H 1967 {\em Phys. Rev. Lett.\/} {\bf 19} 108

\bibitem{Wuphasetransitions}
Lieb E~H and Wu F~Y 1972 Two dimensional ferroelectric models {\em Phase
  Transitions and Critical Phenomena\/} ed Domb C and Green M (Academic Press)

\bibitem{BaxterBook}
Baxter R~J 1982 {\em Exactly Solved Models in Statistical Mechanics\/} (Dover)

\bibitem{Sutherland1967}
Sutherland B 1967 {\em Phys. Rev. Lett.\/} {\bf 19} 103

\bibitem{Korepin1982}
Korepin V 1982 {\em Sov. Phys. Dokl.\/} {\bf 27} 612

\bibitem{KorepinBogoliubovIzergin}
Korepin V~E, Bogoliubov N~M and Izergin A~G 1993 {\em Quantum inverse
  scattering method and correlation functions\/} (Cambridge University Press)

\bibitem{Izergin87}
Izergin A 1987 {\em Sov. Phys. Dokl.\/} {\bf 32} 878

\bibitem{Kuperberg96}
Kuperberg G 1996 {\em Int. Math. Res. Not.\/} {\bf 3} 139

\bibitem{KorepinZinnJustin00}
Korepin V and Zinn-Justin P 2000 {\em J. Phys. A\/} {\bf 33} 7053

\bibitem{ZinnJustin2000}
Zinn-Justin P 2000 {\em Phys. Rev. E\/} {\bf 62} 3411

\bibitem{Syljuasen2004}
Sylju{\aa}sen O and Zvonarev M 2004 {\em Phys. Rev. E\/} {\bf 70} 016118

\bibitem{Allison2005}
Allison D and Reshetikhin N 2005 {\em Annales de l'Institut Fourier\/} {\bf 55}

\bibitem{Reshetikhin}
Reshetikhin N 2008 Lectures on the integrability of the six vertex model {\em
  Exact methods in low dimensional statistical physics and quantum computing\/}
  Les Houches Summer School 89 ed Jacobsen J, Ouvry S, Pasquier V, Serban D and
  Cugliandolo L~F (Oxford University Press)

\bibitem{Bethe35}
Bethe H~A 1935 {\em Proc. Roy. Soc. London A\/} {\bf 150} 552

\bibitem{Pearl}
Pearl J 1988 {\em Probabilistic Reasoning in Intelligent Systems\/} (San
  Francisco, CA, USA: Morgan Kaufmann Publishers Inc.)

\bibitem{Yedidia2003}
Yedidia J~S, Freeman W~T and Weiss Y 2003 Understanding belief propagation and
  its generalizations {\em Exploring Artificial Intelligence in the New
  Millennium\/} ed Lakemeyer G and Nebel B (San Francisco, CA, USA: Morgan
  Kaufmann Publishers Inc.) pp 239--269

\bibitem{CVMreview}
Pelizzola A 2005 {\em Journal of Physics A: Mathematical and General\/} {\bf
  38} R309--R339

\bibitem{MezardMontanari}
M\'{e}zard M and Montanari A 2009 {\em {Information, Physics, and Computation
  (Oxford Graduate Texts)}\/} (Oxford University Press, USA)

\bibitem{Foini2013}
Foini L, Levis D, Tarzia M and Cugliandolo L~F 2013 {\em J. Stat. Mech\/}
  P02026

\bibitem{Levis2013a}
Levis D, Cugliandolo L~F, Foini L and Tarzia M 2013 {\em Phys. Rev. Lett.\/}
  {\bf 110} 207206

\bibitem{Heyderman2013}
Heyderman L~J and Stamps R~L 2013 {\em J. Phys. Cond. Matt.\/} {\bf 25} 363201

\bibitem{NisoliMoessnerSchiffer2013}
Nisoli C, Moessner R and Schiffer P 2013 {\em Rev. Mod. Phys.\/} {\bf 85} 1473

\bibitem{CGP1996a}
Cirillo E, Gonnella G and Pelizzola A 1996 {\em Phys. Rev. E\/} {\bf 53} 1479

\bibitem{CGP1996b}
Cirillo E, Gonnella G and Pelizzola A 1996 {\em Phys. Rev. E\/} {\bf 53} 3253

\bibitem{CGP1997}
Cirillo E, Gonnella G and Pelizzola A 1997 {\em Phys. Rev. E\/} {\bf 55} R17

\bibitem{CGJP1997}
Cirillo E, Gonnella G, Johnston D and Pelizzola A 1997 {\em Phys. Lett. A\/}
  {\bf 226} 59

\bibitem{CGP2000}
Cirillo E, Gonnella G and Pelizzola A 2000 {\em Nucl. Phys. B\/} {\bf 583} 584

\bibitem{CGP2012}
Cirillo E, Gonnella G and Pelizzola A 2012 {\em Nucl. Phys. B\/} {\bf 862} 821

\bibitem{PhysRevB.83.174431}
Phatak C, Petford-Long A~K, Heinonen O, Tanase M and De~Graef M 2011 {\em Phys.
  Rev. B\/} {\bf 83} 174431

\bibitem{PhysRevB.78.144402}
Mengotti E, Heyderman L~J, Fraile~Rodr{\'{\i}}guez A, Bisig A, Le~Guyader L,
  Nolting F and Braun H~B 2008 {\em Phys. Rev. B\/} {\bf 78} 144402

\bibitem{PhysRevLett.111.057204}
Farhan A, Derlet P~M, Kleibert A, Balan A, Chopdekar R~V, Wyss M, Perron J,
  Scholl A, Nolting F and Heyderman L~J 2013 {\em Phys. Rev. Lett.\/} {\bf 111}
  057204

\bibitem{Ladak10}
Ladak S, Read D, Perkins G, Cohen L and Branford W 2010 {\em Nat. Phys.\/} {\bf
  6} 359

\bibitem{ColomoSportiello}
Colomo F and Sportiello A unpublished

\bibitem{Foda2012}
Foda O and Wheeler M 2012 {\em JHEP\/} {\bf 07} 186

\bibitem{Bleher2014}
Bleher P and Liechty K Six-vertex model with partial domain wall boundary
  conditions: ferroelectric phase arXiv:1407.8483

\bibitem{Zinn-Justin2002}
Zinn-Justin P 2002 The influence of boundary conditions in the six-vertex model
  (\textit{Preprint} \eprint{arXiv:cond-mat/0205192})

\bibitem{Budrikis10}
Budrikis Z, Politi P and Stamps R~L 2010 {\em Phys. Rev. Lett.\/} {\bf 105}
  017201

\bibitem{Levis2012}
Levis D and Cugliandolo L~F 2012 {\em EPL\/} {\bf 97} 30002

\bibitem{Budrikis12}
Budrikis Z, Livesey K~L, Morgan J~P, Akerman J, Stein A, Langridge S, Marrows
  C~H and Stamps R~L 2012 {\em New J. Phys.\/} {\bf 14} 035014

\bibitem{Levis2013b}
Levis D and Cugliandolo L~F 2013 {\em Phys. Rev. B\/} {\bf 87} 214302

\bibitem{Nisoli10}
Nisoli C, Li J, Ke X, Garand D, Schiffer P and Crespi V~H 2010 {\em Phys. Rev.
  Lett.\/} {\bf 105} 1

\bibitem{Morgan10}
Morgan J~P, Stein A, Langridge S and Marrows C~H 2011 {\em Nature Phys.\/} {\bf
  7} 75

\bibitem{Morgan13}
Morgan J~P, Akerman J, Stein A, Phatak C, Evans R~M~L, Langridge S and Marrows
  C~H 2013 {\em Phys. Rev. B\/} {\bf 87} 024405

\end{thebibliography}

\end{document}